\documentclass{PoS}

\usepackage{amsmath}
\usepackage{amsfonts,amscd,dsfont}
\usepackage{amssymb}
\usepackage{float}
\usepackage{cite}
	
\newcommand{\cN}{\mathcal{N}}
\newcommand{\cG}{\mathcal{G}}
\newcommand{\cF}{\mathcal{F}}
\newcommand{\cR}{\mathcal{R}}
\newcommand{\bp}{\mathbf{p}}
\newcommand{\bq}{\mathbf{q}}
\newcommand{\bx}{\mathbf{x}}
\newcommand{\im}{\text{i}}
\newcommand{\diff}{\text{d}}
\newcommand{\zbar}{\bar{z}}

\title{Multi-Loop Amplitudes in the High-Energy Limit in N=4 SYM}

\ShortTitle{MRK in N=4 SYM}

\author{Vittorio Del Duca,$^a$ Stefan Druc,$^b$ James Drummond,$^b$ Claude Duhr,$^{cd}$ Falko Dulat,$^e$ \speaker{Robin Marzucca},$^{d}$ Georgios Papathanasiou$^f$ and Bram Verbeek$^d$ \\
\llap{$^a$} Institute for Theoretical Physics, ETH Z\"urich, Wolfgang-Pauli-Stra{\ss}e 27, 8093 Z\"urich, Switzerland\\
\llap{$^b$} School of Physics \& Astronomy, University of Southampton, Highfield, Southampton, SO17 1BJ, United Kingdom\\
\llap{$^c$} Theoretical Physics Department, CERN, CH-1211 Geneva 23, Switzerland\\
\llap{$^d$} Centre for Cosmology, Particle Physics and Phenomenology (CP3),\\
					Universit\'e catholique de Louvain, Chemin du Cyclotron 2, 1348 Louvain-La-Neuve, Belgium\\
\llap{$^e$} SLAC National Accelerator Laboratory, Stanford University, 2575 Sand Hill Road, Menlo Park, Stanford, CA 94309, USA\\
\llap{$^f$} DESY Theory Group, DESY Hamburg, Notkestra{\ss}e 85, D-22607 Hamburg, Germany\\

E-mail: \email{delducav@itp.phys.ethz.ch},
			\email{sd3g14@soton.ac.uk}, 
			\email{J.M.Drummond@soton.ac.uk},
			\email{claude.duhr@cern.ch},
			\email{dulatf@slac.stanford.edu},
			\email{robin.marzucca@uclouvain.be},
			\email{georgios.papathanasiou@desy.de},
			\email{bram.verbeek@uclouvain.be}}

\abstract{We introduce a novel way to perform high-order computations in multi-Regge-kinematics in planar $\cN = 4$ supersymmetric Yang-Mills theory and generalize the existing factorization into building blocks at two loops to all loop orders. Afterwards, we will explain how this framework can be used to easily obtain higher-loop amplitudes from existing amplitudes and how to relate them to amplitudes with higher number of legs.}

\FullConference{Loops and Legs in Quantum Field Theory (LL2018)\\
		29 April 2018 - 04 May 2018\\
		St. Goar, Germany}

\begin{document}

\section{Introduction}

Vast progress in understanding the structure of the S-Matrix in $\cN = 4$ Super Yang-Mills (SYM) theory has been made in recent years. Large part of this success has been due to the very large amount of symmetries and special properties of the theory, which make it a prime candidate for the search for general mathematical structures of gauge theory scattering amplitudes. In the planar limit, the conformal symmetry of the theory closes with dual conformal symmetry to form an infinite-dimensional Yangian-symmetry \cite{DualConformal1,DualConformal2,DualConformal3,Yangian}, which is often seen to be a criterion for integrability. Since we currently do not know an all-order solution of planar $\cN = 4$ SYM, we would like to know the scattering amplitudes in as many kinematical regimes as possible. One of them is a very special kinematical limit, called the multi-Regge-limit, in which it was possible to compute amplitudes with many external legs to very high orders in perturbation theory.

\section{The Remainder Function in MRK}

\begin{figure}[!h] 
  \centering
  \includegraphics[width=0.4\textwidth]{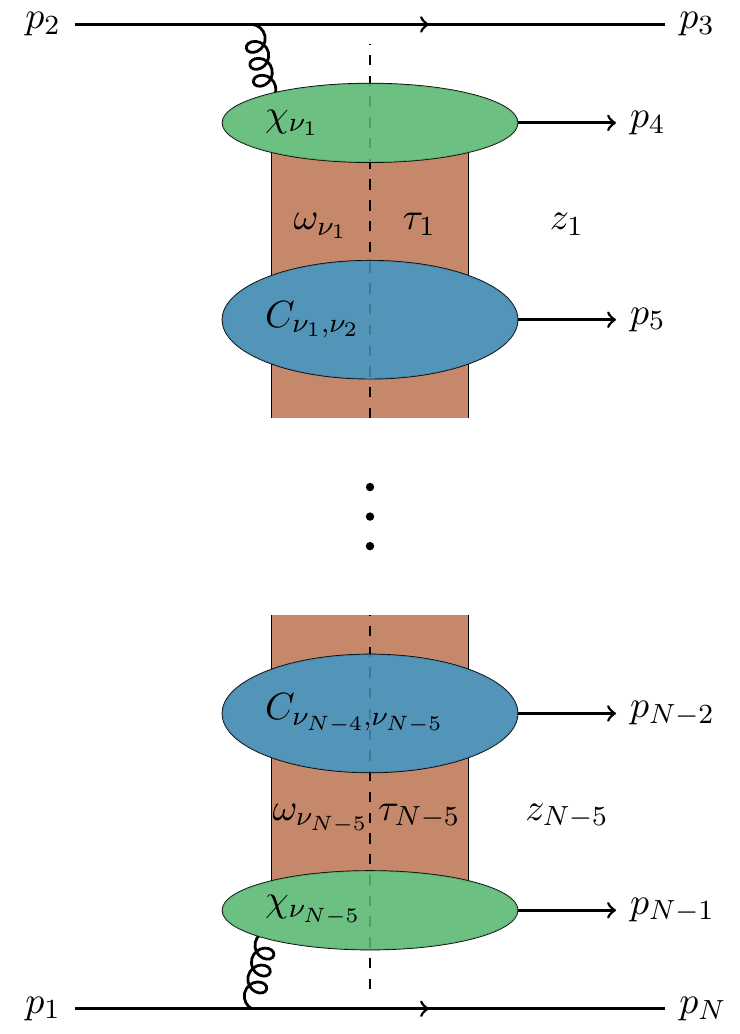}
  \caption{Visualization of BFKL factorization in Fourier-Mellin space.}
  \label{fig:FMFactorization}
\end{figure}

The dual conformal symmetry of planar $\cN = 4$ SYM is restrictive enough to fix all four- and five-point amplitudes to all loop orders \cite{BDS}. It is only at six points that we see for the first time the appearance of the remainder function $R_N$ or the BDS normalized ratio $\cR_N$, respectively \cite{DualConformal2,DualConformal3}, and we have

\begin{align}
A_N =\left\lbrace \begin{array}{ll}
A_N^{\text{BDS}} e^{R_N}, & \text{ MHV}\\
A_N^{\text{BDS}} \cR_N, & \text{ otherwise},
\end{array}  \right.
\end{align}
where $A_N^{\text{BDS}}$ is the so-called BDS ansatz \cite{BDS}, which describes the amplitude exactly to all orders for 4 and 5 external legs.
We will consider the color ordered scattering amplitude of $N$ gluons with all external momenta outgoing. Let us first define lightcone and complex transverse coordinates as
\begin{align}
p^{\pm} & \equiv p^0 \pm p^z, & \bp & \equiv p_{\perp} = p^x + \im p^y.
\end{align}
Then the scalar product of two vectors $p,q$ is given by
\begin{equation}
2 p\cdot q = p^+ q^- + p^- q^+ - \bp \bar{\bq} - \bar{\bp} \bq.
\end{equation}
Let us further, without loss of generality, choose a reference frame such that the momenta $p_1, p_2$ of the two initial state gluons are aligned with the $z$-axis with $p_2^z = p_2^0$, which yields $p_1^+ = p_2^- = \bp_1 = \bp_2 = 0$. Then the multi-Regge-limit \cite{Regge} is defined as the limit where the remaining external momenta $p_i$, $3 \leq i \leq N$ are strongly ordered in rapidity while having no hierarchy in the transverse components, or equivalently 

\begin{align}
p_3^+ \gg 
\dots 
\gg p_N^+, \hspace{1cm} |\bp_3| \simeq \dots \simeq |\bp_N|.
\end{align}
As the initial gluons are barely deflected in this limit, their helicity must be conserved along their path, so that the amplitude will only depend on the helicities $h_1, \dots, h_{N-4}$ of the produced gluons. We will therefore often label the BDS normalized ratio $\cR$ only with the helicities of the produced gluons

\begin{equation}
\cR_{h_1,\dots,h_{N-4}} = \frac{A_N(-,+,h_1,\dots,h_{N-4},+,-)}{A_N^{\text{BDS}}(-,+,h_1,\dots,h_{N-4},+,-)}.
\end{equation}
The ratio $\cR_{h_1,\dots,h_{N-4}} = 1$ in MRK in the Euclidean region. After analytically continuing the energy components of the produced gluons, however, we find a non-trivial expression. Due to fixing a hierarchy in the longitudinal component of the external momenta, the amplitude exhibits logarithms $\log \tau_i $, with 
\begin{equation}
\tau_i = \delta_i \sqrt{\frac{|\bq_{i-1}|^2|\bq_{i+1}|^2|\bp_{i+3}|^2}{|\bq_{i}|^4 |\bp_{i+4}|^2}},
\end{equation}
where $\bq_i = \bx_{i+2} - \bx_1$, $\delta_i = p_{i+4}^+ / p_{i+3}^+ \rightarrow 0$, and where we define the dual coordinates $\bx_i$ via 
\begin{equation}
\bp_{i+3} = \bx_{i+2}-\bx_{i+1}, \hspace{1cm} i=0, \dots , N-4.
\end{equation}
These logarithms become very large as we approach the multi-Regge-limit and should therefore be resummed, which yields 

\begin{equation}
\begin{split}\label{eq:MRK_conjecture}
\cR_{h_1\dots h_{N-4}}= &1  + a\,i\pi\, \left[ \prod_{k=1}^{N-5}\sum_{n_k=-\infty}^{+\infty}\left(\frac{z_k}{\zbar_k}\right)^{\frac{n_k}{2}}\int_{-\infty}^{+\infty}\frac{d\nu_k}{2\pi}|z_k|^{2i\nu_k} \right]
 \\ 
 &\times
 \left[\prod_{k=1}^{N-5}e^{-L_k \omega_k}\right]\, \chi^{h_1}_{1} \left[\prod_{k=2}^{N-5}C^{h_k}_{k-1,k} \right] \,\chi^{-h_{N-4}}_{N-5},
\end{split}
\end{equation}
where $\omega_k, \chi_k^\pm$ and $C^\pm_{ij}$ are the BFKL eigenvalue, impact factor and central emission blocks respectively and $L_k = \log \tau_k + \im \pi$. This is in accordance with the conjectured to factorization of the remainder function in an auxillary space, also referred to as Fourier-Mellin space, to all orders in perturbation theory into a small number of building blocks \cite{ReggeCut,BFKL6pt,BFKL7pt} which can be visualized as in fig \ref{fig:FMFactorization}. The initial gluons are connected via impact factors $\chi$ to a radiated gluon and to a ladder of Reggeon exchanges and central emission blocks $C$, with an additional gluon radiated from each central emission block and a large logarithm $\log \tau_k $ for each reggeized gluon. 
Expanding the result perturbatively, we find at every loop order a polynomial in these large logarithms. Truncating this expansion at the highest appearing order of logarithms, we get the leading logarithmic approximation (LLA) of the amplitude. For the remainder of this paper, we will focus on exactly this approximation and comment on the extension beyond LLA in the end. The corresponding expression for the LLA remainder function can be obtained by replacing the building blocks in \eqref{eq:MRK_conjecture} by their respective leading terms
\begin{align}
\omega_k &\rightarrow -a E_{k} & \chi^\pm_{k} &\rightarrow \chi_{0,k}^\pm  & C^\pm_{ij} &\rightarrow C_{0,ij}^\pm
\end{align} 
Then the $\ell$-loop $N$-point remainder function takes the form 

\begin{align} \label{eq:intro_g}
R^{(\ell)}_N = 2 \pi \im  a^\ell \sum_{i_1 + \dots + i_{N-5} = \ell-1} \left( \prod_{k=1}^{N-5} \frac{\log^{i_k} \tau_k}{i_k !} \right) g_{h_1 \dots h_{N-4}}^{(\ell,i_1,\dots,i_{N-5})}(z_1,\dots,z_{N-5}).
\end{align}
Further, since the logarithms $\log \tau_k$ capture the dependence of the amplitude on the longitudinal part of the external momenta, the \emph{perturbative coefficients} $g_{h_1\dots h_{N-4}}^{(\ell;i_1,\dots,i_{N-5})}$ depend only on the $N-2$ transverse momenta $\bp_3 , \dots , \bp_N$. Let us further define the cross-ratios 
\begin{equation}
z_i = \frac{(\bx_1 - \bx_{i+3})(\bx_{i+2}-\bx_{i+1})}{(\bx_1 - \bx_{i+1})(\bx_{i+2}-\bx_{i+3})}
\end{equation}
for later convenience. \\  \\
It is expected that amplitudes in MRK in $\cN = 4$ SYM can be expressed in terms of polylogarithms $G_{a_1,\dots,a_n}(z)$ \cite{Brown,Ours1}. These are defined via the recursion

\begin{equation}
G_{a_1,\dots,a_n}(z) = \int_{0}^{z} \diff z \, \frac{1}{z-a_1} G_{a_2,\dots,a_n}(z) \hspace{1cm} G(z) = 1
\end{equation}
and 
\begin{equation}
G_{\underbrace{0,\dots,0}_{n \text{ times}}}(z) = \frac{1}{n!}\log^n (z).
\end{equation}
From the optical theorem, we can further learn about possible branch cuts of the amplitudes, namely they originate from points at which virtual particles become on-shell. Since all particles in $\cN = 4$ SYM are massless, branch-cuts need to start at $(x_i - x_j)^2 = 0$. Going to MRK, this condition translates to  branch-cuts starting at $|\bx_i - \bx_j|^2= 0$. Since $|\bx_i - \bx_j|^2 \geq 0$, this means that the perturbative coefficients are single-valued functions. A more rigorous derivation based on symbols and cluster algebras can be found in \cite{SV,Ours1}. At this point it is convenient to introduce a basis of single-valued functions to express the perturbative coefficients in. The so called \emph{single-valued polylogarithms} $\cG_{a_1,\dots,a_n}(z)$ \cite{Brown} are linear combinations of polylogarithms in $z$ and $\zbar$, with $\zbar$ being the complex conjugate of $z$, such that all branch-cuts cancel and they fulfil the same holomorphic differential equation as regular polylogarithms,

\begin{equation}
\partial_z \cG_{a_1,\dots,a_n}(z) = \frac{1}{z-a_1}\cG_{a_2,\dots,a_n}(z).
\end{equation}
%
%
%

\section{The Perturbative Coefficients}

\begin{figure}[!h] 
  \centering
  \includegraphics[width=0.6\textwidth]{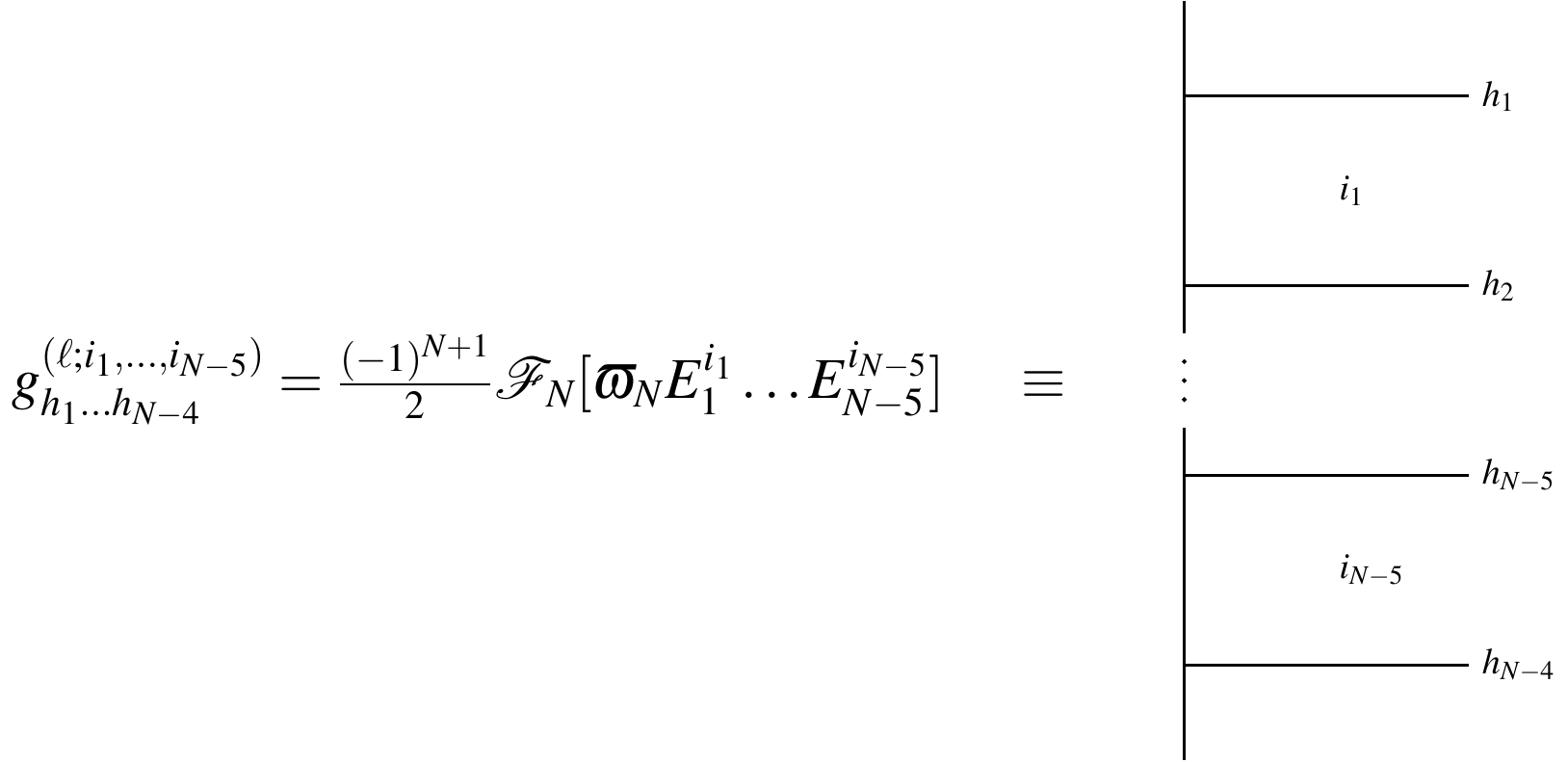}
  \caption{Graphical representation of perturbative coefficients.}
  \label{fig:LadderIntro}
\end{figure}

Expanding \eqref{eq:MRK_conjecture} in the coupling constant and comparing coefficients with \eqref{eq:intro_g}, we can read off the definition of the $g_{h_1 \dots h_{N-4}}^{(\ell,i_1,\dots,i_{N-5})}$ and we find at LLA, i.e. for $\sum i_k = \ell-1$,
\begin{align}
g_{h_1 \dots h_{N-4}}^{(\ell,i_1,\dots,i_{N-5})} &= \frac{(-1)^{N+1}}{2} \left[ \prod_{k=1}^{N-5}\sum_{n_k=-\infty}^{+\infty}\left(\frac{z_k}{\zbar_k}\right)^{\frac{n_k}{2}}\int_{-\infty}^{+\infty}\frac{d\nu_k}{2\pi}|z_k|^{2i\nu_k} \right] \varpi_N E_1^{i_1} \dots E_{N-5}^{i_{N-5}} \\
	&\equiv \frac{(-1)^{N+1}}{2} \cF_N[ \varpi_N E_1^{i_1} \dots E_{N-5}^{i_{N-5}} ], \label{eq:pertCoeff}
\end{align}
where 
\begin{equation}
\varpi_N = \chi_{0,1}^{h_1} C_{0,12}^{h_2} \dots C_{0,(N-6) (N-5)}^{h_{N-5}} \chi_{0,N-5}^{-h_N-4}
\end{equation}
is the \emph{vacuum ladder}. Its name will become clear, after introducing a graphical representation for the perturbative coefficients $g_{h_1 \dots h_{N-4}}^{(\ell,i_1,\dots,i_{N-5})}$. At LLA, the expansion of \eqref{eq:MRK_conjecture} is fixed up to how far we expand the individual exponentials $\tau_k^{a E_k}$. We will therefore view the the leading term, i.e. with all $\tau_k^{a E_k} \rightarrow 1$, as a vacuum state and occurrences of $E_k$, coming from higher orders of the exponentials as insertions. We can symbolize the vacuum state as a ladder of external gluons and insertions of $E_k$ as insertions into the faces of the diagram, as can be seen in fig. \ref{fig:LadderIntro}.

Now that we have reviewed the structure of the remainder function in MRK in $\cN = 4$ SYM, as well as its mathematical properties, we want to focus on how to compute the perturbative coefficients in the next section.

\section{Fourier-Mellin Convolutions}

In this section, we will explain how to use convolutions to easily compute scattering amplitudes in MRK. For this purpose, we will restrict ourselves to the MHV case for now and extend the concept to the non-MHV case afterwards. As we have seen in \eqref{eq:MRK_conjecture}, the remainder function $R_N$ corresponds to a multiple Fourier-Mellin transform
\begin{equation}
\cF [F(\nu,n) ] = \sum_{n=-\infty}^\infty \int \frac{\diff \nu}{2 \pi} \left( \frac{z}{\zbar}\right)^{\frac{n}{2}} |z|^{2 \im \nu} F(\nu,n).
\end{equation}
Under this transformation, products are mapped onto convolutions, and we have

\begin{equation}
\cF[F \cdot G] = \cF[F] * \cF[G] = \frac{1}{\pi} \int \frac{\diff^2 w}{|w|^2} \cF[F] (w) \cF[G] \left(\frac{z}{w} \right).
\end{equation}
This relation allows us to relate perturbative coefficients of different loop orders by repeatedly extracting leading order BFKL eigenvalues $E_i$ from the Fourier-Mellin integral \eqref{eq:pertCoeff}, allowing us to raise the loop order of a given perturbative coefficient by computing a convolution integral. At seven points, for example, we have

\begin{equation}
g_{+++}^{(\ell;i_1,i_2)} = g_{+++}^{(\ell-1;i_1-1,i_2)} * \cF[E_1] = g_{+++}^{(1;0,0)} * \cF[E_1]^{*i_1} * \cF[E_2] ^{*i_2}.
\end{equation}
The necessary ingredient, namely the Fourier-Mellin transform of the leading order BFKL eigenvalue,
\begin{equation}
\cF[E_k] = - \frac{z_k + \zbar_k}{2 |1-z_k|^2},
\end{equation}
evaluates to a simple rational function. In addition to this, we will exploit the fact that the perturbative coefficients are single-valued objects. It was shown that the integral over the whole complex plain of a single-valued function $f$ can be calculated by computing the holomorphic residues of a single-valued anti-holomorphic primitive $F$ of $f$ \cite{Schnetz},

\begin{equation}
\int \frac{\diff^2 z}{\pi} f(z) = \text{Res}_{z=\infty} F(z) - \sum_i \text{Res}_{z=a_i}F(z), \hspace{1cm} \partial_{\zbar} F(z) = f(z), 
\end{equation}
where $\lbrace a_i,\infty \rbrace$ is the set of singularities of $F$. Close to any of these points, a single-valued function $F(z)$ can be expanded as
\begin{align}
F(z) &= \sum_{k,m,n} c_{k,m,n}^{a_i} \log  \left|1-\frac{z}{a_i} \right|^2 (z-a_i)^m (\zbar - \bar{a}_i)^n, & z &\rightarrow a_i \\
F(z) &= \sum_{k,m,n} c_{k,m,n}^{\infty} \log  \frac{1}{|z|^2} \frac{1}{z^m} \frac{1}{\zbar^n}, & z &\rightarrow \infty.
\end{align}
Then the holomorphic residue is defined as 
\begin{equation}
\text{Res}_{z=a}F(z) \equiv c_{0,-1,0}^a.
\end{equation}
With this at hand, once we know a starting point, e.g. the two-loop remainder function, we can promote it to higher loops simply by convoluting it with $\cF[E_k]$, which corresponds to a rather straight-forward computation of residues.

\section{Factorization}

Expressing the perturbative coefficients in terms of the dual coordinates $\bx_i$ instead of the cross ratios $z_i$, we find that in certain cases the dependence on some of them drops out and we are left with substantially simpler objects. In particular, it turns out that any perturbative coefficient with empty faces and equal neighbouring helicities will simplify to a lower-point object \cite{Ours1}. The eight-point perturbative coefficient $g_{h_1 h \; h \; h_2}^{(5;3,0,1)}$ for example reduces to the seven-point perturbative coefficient $g_{h_1 h \; h_2}^{(5;3,1)}$.

\begin{figure}[H] \label{fig:Factorization8pt}
  \centering
  \includegraphics[width=0.4\textwidth]{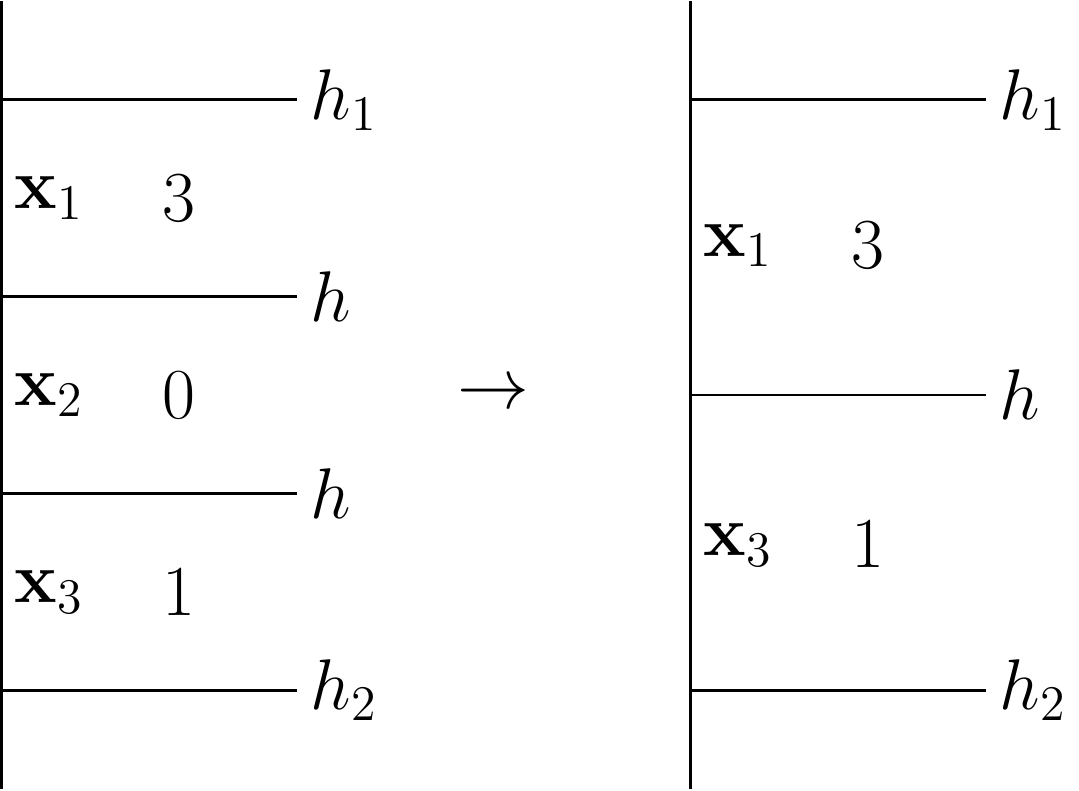}
\end{figure}

Since the only insertions we can have at LLA are leading order BFKL eigenvalues $E_k$ and since at MHV all helicities are equal, we can conclude that at any given loop-order, the remainder function $R_N$ can be expressed in terms of a finite number of perturbative coefficients. At three loops, for example we find \\

\begin{equation}
R_{+\dots+}^{(3)} = \sum_{1\leq i \leq N-5} \frac{1}{2} \log \tau_i^2 g_{++}^{(3;2)}(\bx_i) + \sum_{1 \leq i < j \leq N-5}\log \tau_i \log \tau_j g_{+++}^{(3;1,1)}(\bx_i,\bx_j),
\end{equation}
where we decided to only label the remainder function by its helicity configuration and the number of external legs is implicit.

\section{Amplitudes Beyond MHV and LLA}

Fourier-Mellin convolutions can also be used to flip the helicities of external particles. At seven points, for example we have 
\begin{equation}
\cF[\chi_1^- C_{12}^+\chi_2^-] = \cF \left[ \frac{\chi_1^-}{\chi_1^+} \right] * \cF[\chi_1^+ C_{12}^+\chi_2^-],
\end{equation}
with
\begin{equation}
\cF \left[ \frac{\chi_1^-}{\chi_1^+} \right]  = - \frac{z_1}{(1-z_1)^2}.
\end{equation}
We see that the relevant integration kernel for the convolutions is again a simple rational function. Note that the condition for a perturbative coefficient to simplify to a lower-point coefficient included that the face with no insertions was bounded by two external lines with equal helicities. This means that some amplitudes beyond MHV, e.g. the amplitude with alternating helicities, will not be reducible to lower-point objects. Most of them, like for example $R^{(2)}_{-+\dots+}$ will however still simplify drastically, like
\begin{equation}
R^{(2)}_{-+\dots+} = \log \tau_1 g_{-+}^{(2;1)}(\bx_1) + \sum_{1\leq i \leq N-5} \log \tau_i g_{-++}^{(2;0,1)}(\bx_1,\bx_i).
\end{equation}\\ \\
Beyond LLA, we will for the first time encounter corrections to the BFKL building blocks in \eqref{eq:MRK_conjecture}, the individual perturbative coefficients will however still correspond to Fourier-Mellin transformations and hence they can be computed from a starting point using convolutions \cite{Ours2}. We will however encounter additional insertions, both into the faces as well as vertices of our graphical representation. \\ \\
We have applied the described mathematical framework both at LLA and at NLLA. At LLA, we have computed all MHV amplitudes through 5 loops, and all 8 point amplitudes through 4 loops in all helicity configurations. At NLLA, we have computed all MHV 3 loop amplitudes and the 7 point amplitude in the MHV configuration through 5 loops and in the $+-+$ and $-++$ helicity configurations through 3 and 4 loops, respectively.

\section{Conclusions}

We have presented a framework with which scattering amplitudes in MRK in $\cN = 4$ SYM can easily be computed to high orders in perturbation theory and for many external particles. In particular it is possible for the first time, at any order in perturbation theory, to compute an infinite number of amplitudes, albeit in a special kinematical limit. By calculating a finite number of perturbative coefficients, we can immediately write down the remainder function for any number of external particles. The framework is applicable for all helicity configurations as well as at LLA and NLLA and it allowed us to completely classify the function space of scattering amplitudes within these accuracies.  \\ \\
Beyond NLLA, we do not expect any conceptually new hurdles. Since the central emission block , as opposed to the impact factors and the BFKL eigenvalue, is only known to NLO, we have however not been able to test our framework there. It would certainly be interesting to check explicitly whether it holds to all logarithmic accuracies and whether it is possible to determine the function space of MRK to all orders.

\section{Acknowledgements}

This work is supported by the European Research Council (ERC) through the grants 637019 (MathAm) and 648630 (IQFT), and by the U.S. Department of Energy (DOE) under contract  DE-AC02-76SF00515.

\end{document}